\documentclass[aps, superscriptaddress, showkeys]{revtex4}
\usepackage{amsmath,amssymb,graphicx}
\begin{document}
\title{Interpreting Electrical-Resistivity Tomography measurements using Neural Network}
\date{\today}
\author{Itay Naeh}
\thanks{E-mail: itay@naeh.us, yitzhak.peleg@gmail.com}
\author{Yitzhak Peleg}
\thanks{I. N. and Y. P. contributed equally to this work}
\affiliation{Rafael Advanced Defense Systems Ltd.}
\author{Alex Furman}
\affiliation{Technion - Israel Institute of Technology, Civil and Environmental Engineering, Haifa 32000, Israel}
\author{Shie Mannor}
\affiliation{Technion - Israel Institute of Technology, Faculty of Electrical Engineering, Haifa 32000, Israel}

\begin{abstract}
Electrical Resistivity Tomography (ERT) has been extensively used for  imaging the subsurface resistivity distribution and structure.
Over the years, many algorithms have been developed in order to solve the subsurface resistivity distribution from the ERT measurements.
In this paper a new method for interpreting the ERT measurements is presented. Using supervised learning to train a neural network, we are able to interpret the ERT measurement into a 2D image of the underground resistivity up to depths of 50 meters while using a simple Wenner-Schlumberger survey of 96 electrodes with 1 meter spacing. The neural network is trained and tested using simulative data and it is shown to have superior results over a well established inversion method.

\end{abstract}

\keywords{Neural Networks; Electric Resistivity Tomography}
\maketitle

\section{Introduction}\label{sec: Intro}
Electrical Resistivity Tomography (ERT) is a method which allows the imaging of the electrical conductivity profile of the ground in an unobtrusive manner \cite{morelli1996advances}. A typical survey, as investigated in this work, is conducted by placing electrodes in a linear formation, with some spacing $d$ between the electrodes.
Measurements are performed by applying a near steady-state current (up to a few amperes) between two electrodes and measuring the voltage between two other electrodes. This procedure is carried out using all the relevant electrodes for the survey \cite{saad2012groundwater}.

Due to the fact that the current flows in an unknown path which is determined by the conductivity profile of the ground, the interpretation is not straightforward. Moreover, the number of unknown in the problem is larger than the number of equations; therefore this problem is ill-posed and needs to be solved using iterative inversion methods \cite{tsourlos1999algorithm, WRCR21447}.
The inversion process begins with an initial (hopefully intelligent) guess of the under-surface resistivity and the forward simulated measurements on this guessed profile.
At each iteration the resistivity profile is changed in a certain way in order to minimize some cost function (which includes the difference between the simulated forward solutions and the actual measurements and some other constrains) \cite{WRCR10223, loke2014smoothness}. 
However, the final solution is sensitive to both the initial guess and the choice of constrains and cost function. In order to overcome those difficulties in interpreting the ERT measurements, providing a 2D imaging of the subsurface resistivity profile, this paper introduces a different approach based on supervised learning for training a Neural Network (NN).

In the past few years there have been a major advancement in training and implementing neural networks (NN) in many different fields. Nowadays, (deep) Neural Networks are the state-of-the-art in many classification and prediction tasks.
NN are now able to translate sentences \cite{NN-Translate}, generate new art \cite{NN-GAN} and even define communication protocol \cite{NN-Oshri}.
The idea of using a NN for interpretation of the under-surface geophysical properties is not new and has been suggested almost three decades ago \cite{NN1991}.
During recent years NN have been applied to analyzing gravity measurments \cite{NN-gravity}, seismic waveform inversion \cite{NN-seismic}, 
data from ground penetrating radar \cite{NN-GPR} and more. 

A prior work was performed by \cite{NN-VES} which demonstrated the ability to interpret geophysical measurements by artificial neural networks. The assumption of the solution was of a layered ground profile and the solution determined the depths and resistivities of 5-6 layers. In our case the output is a full 2D image without the need to define, a-priori, a sub-surface model. Another major difference is the scales, where in \cite{NN-VES} the survey had hundreds of meters of electrode separation and the depths of interpretation were 1 km, trading off resolution. However, we aim our work at shallow ERT and high resolution of $1m^2$.

The rest of this manuscript is organized as follows.
The definition of the ground resistivity profile and details about modelling the ERT survey are given in section \ref{sec: Model}.
Section \ref{sec: NN} deals with the description of the NN and the database we used for training it.
The results of the NN and a comparison to a widely used inversion program are shown in section \ref{sec: Results} and concluding remarks are given at section \ref{sec: Conclusions}.

\section{Model and Simulations}\label{sec: Model}
\begin{figure}[t]
	\includegraphics[scale=0.5]{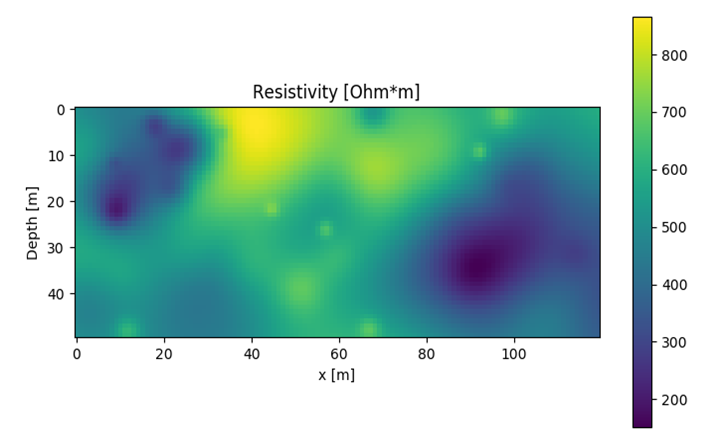}
	\caption{A representative sample of the ground used for training and testing the neural network.}
	\label{fig: ground}
\end{figure}

The subsurface structure is (in general) inhomogeneous thus, in order to increase the validity of the presented method, we have simulated inhomogeneous ground by assuming random resistivity values with spatial correlations. The simulated underground structure is a matrix $G \in R^{120 \times 50}$ where each element, $G_{i,j}$, represents the resistivity of a $1m^2$ cross section at horizontal position $i$ and depth $j$.
The ground resistivity profile is a linear gradient perturbed by $N=100$ random $2D$ Gaussian
\begin{equation}\label{eq: ground}
\rho = 600 - 0.2z + \sum_{n=1}^{N} A_n \mathcal{N}(\mu_n, \sigma_n)
\end{equation}
where $z$ is measured in meters, $\mathcal{N}(\mu, \sigma)$ is a $2D$ Gaussian centred at $\mu$ with $\sigma$ width. The values of $A_n, \sigma_n$ are taken from uniform distributions $U(-100, 100)$ and $U(3, 30)$, respectively. The center of the Gaussians,
$\mu_n$ are random variables uniformly distributed in the underground. A typical simulated ground is depicted in Figure \ref{fig: ground}.

\begin{figure}[t]
	\includegraphics[width=\textwidth, height=5cm]{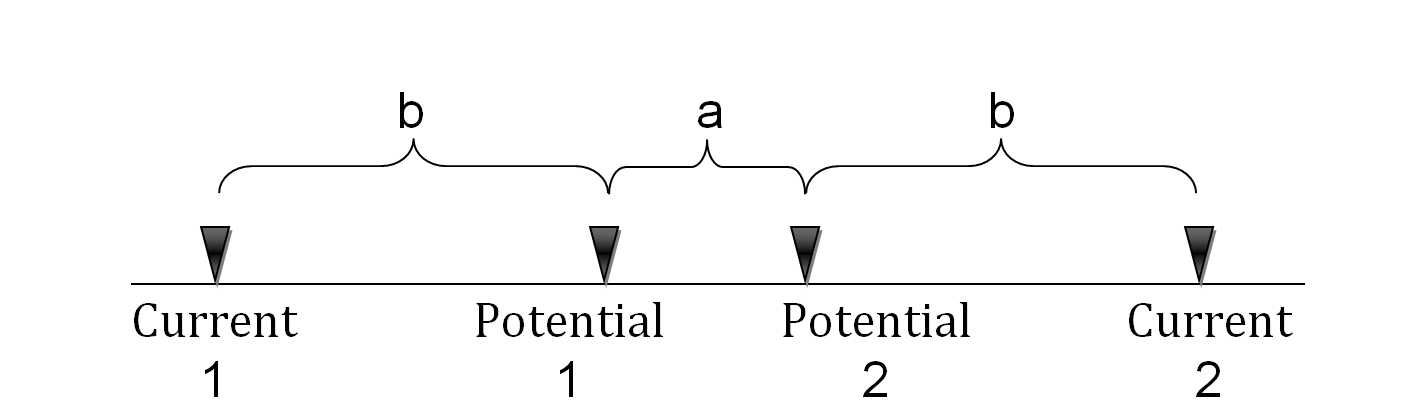}
	\caption{Description of the Wenner-Schlumberger survey. $b=na$ where $n\in\left\{1, 2\right\}$}
	\label{fig: survey}
\end{figure}

There are several well-known survey protocols that are commonly used today. The main differences between the protocols are: 
(i) the  depth  of  investigation, 
(ii)  the  sensitivity  of  the  array to vertical and horizontal changes in the subsurface resistivity, 
(iii) the horizontal data coverage and 
(iv) the signal strength \cite{loke-tutorial}.
Since the ground model we used in the simulations have changes of resistivity values in both horizontal and vertical directions, 
we have decided to use the “Wenner-Schlumberger” (WS) survey method.
The main advantage of the WS array is its sensitivity to both vertical and horizontal changes. 

The WS array is defined by three parameters: $a$ which is the distance between the inner potential electrodes and $b=na \ (n\in\mathcal{Z})$ which defines the distance between a potential electrode to the closest current electrode in terms of multiplication of $a$ as shown in Fig. \ref{fig: survey}. A third parameter of the array is the location of the left current electrode (\textit{current 1} in fig. \ref{fig: survey}). In order to maintain a reasonable number of measurements, we used $n\in\left\{1, 2\right\}$ and $a\in\left\{1, 2, \ldots ,33\right\}$ and for each value of the pair $(n, a)$ we used all possible position for the \textit{current 1} electrode. As a result, this protocol contained 1488 measurements of the apparent resistivity. The forward simulation of the  ERT measurements was carried out using R2 software \cite{R2} and the simulated ground has been extended far away to minimize boundary effects on the results.

\section{Neural Network}\label{sec: NN}
When trying to interpret the measurements of an ERT survey (as in any inversion problem) one asks the question what is the medium which caused the acquired measurements. Thus, if we have enough examples of the medium and the out-coming ERT measurements we can try and get insights about the inverse relation between the measurements and the medium which caused them. This approach, for training a neural network is called "supervised learning".

We have simulated a massive database of $15,000$ examples - random grounds as defined by \eqref{eq: ground} and a forward modelling of the 1488 ERT measurements has been carried out for each ground. The database thus consisted of pairs $(G_k, E_k)$ - the ground matrix and the resulting ERT measurements vector, respectively. We have randomly split the dataset into a training set consisting of approximately $90\%$ of the examples and the remaining $1500$ examples were used to test the training of the neural network.

The input of the neural network is the ERT measurements vector $E$ which consist of 1488 measurements and the output is a ground matrix with dimensions $120 \times 50$. In order to have as simple solution as possible, we used only one hidden layer of 250 neurons with a $ReLU$ activation function
\begin{equation}\label{eq: ReLU}
ReLU(x) = 
\begin{cases}
	0 \quad x<0 \\
	x \quad x>=0
\end{cases}
\end{equation}
The overall model is given by
\begin{equation}\label{eq: model}
\widetilde{\textbf{g}} = \textbf{b}^{(1)} + W^{(1)}_{6000 \times 250} 
\text{ReLU}
\left(\textbf{b}^{(0)} + W^{(0)}_{250 \times 1488} \textbf{E} \right) \end{equation}
where \textbf{bold} letters stand for vectors. $W^{(0)}, W^{(1)})$ are the weight matrices of the corresponding layers and $\textbf{b}^{(0)}, \textbf{b}^{(1)}$ are the corresponding bias vectors. 
The output vector $\widetilde{\textbf{g}}$ is reshaped into a matrix $\widetilde{G}_{120 \times 50}$ which is the ground predicted by the neural network model.

In order to train the NN, a loss (a.k.a. cost) function must be defined. In many application one uses the euclidean distance between the prediction $\widetilde{G}$ and ground truth $G$. However, in this scenario it turned out that an $L_1$ loss function
\begin{equation}\label{eq: L1}
L_1 = \left\| \widetilde{G} - G \right\|_1 = \sum_{i,j} \text{abs}\left(\widetilde{G}_{i,j} - G_{i,j}\right)
\end{equation}
results in a better performance and shorter convergence time.

The weights and biases were initializes to be normally distributed with standard deviation of $1E-6$ and we used the Adam optimizer to train the NN. Each batch consisted of 50 examples pairs of ground and ERT measurements $(G, \textbf{E})$ and the network has been trained for 1000 epochs.

\section{Results and comparison}\label{sec: Results}

\begin{figure}
	\includegraphics[width=\textwidth]{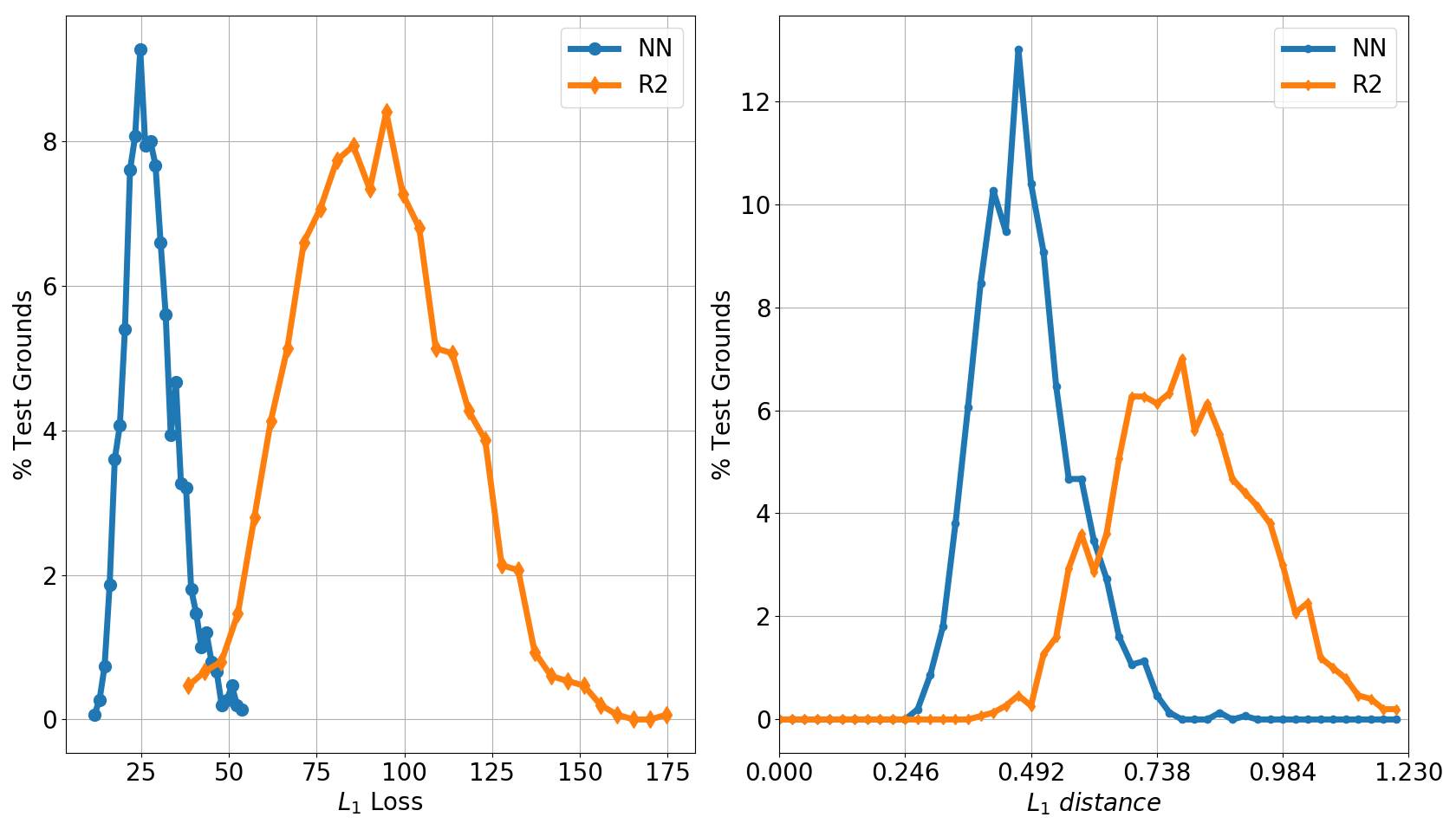}
	\caption{
	\textbf{LEFT PANEL}: Histograms of $L_1$ loss of the predictions of the neural network and the predictions of the classical inversion method by using R2. In almost all test cases the $L_1$ loss of the NN predictions is smaller than the $L_1$ loss of the inversion method.
	\textbf{RIGHT PANEL}: Histograms of $L_1$ histograms distance between the predictions of the neural network and the predictions of R2 with respect to the true
grounds. The $L_1$ distance of the histograms of the prediction made by the neural network, in most cases, is smaller than $L_1$ distance of the histograms of the inversion method (R2).}
	\label{fig: l1_and_l1h} 
\end{figure}

\begin{table}[!t]
	\begin{tabular}{c}
		\includegraphics[scale=0.4]{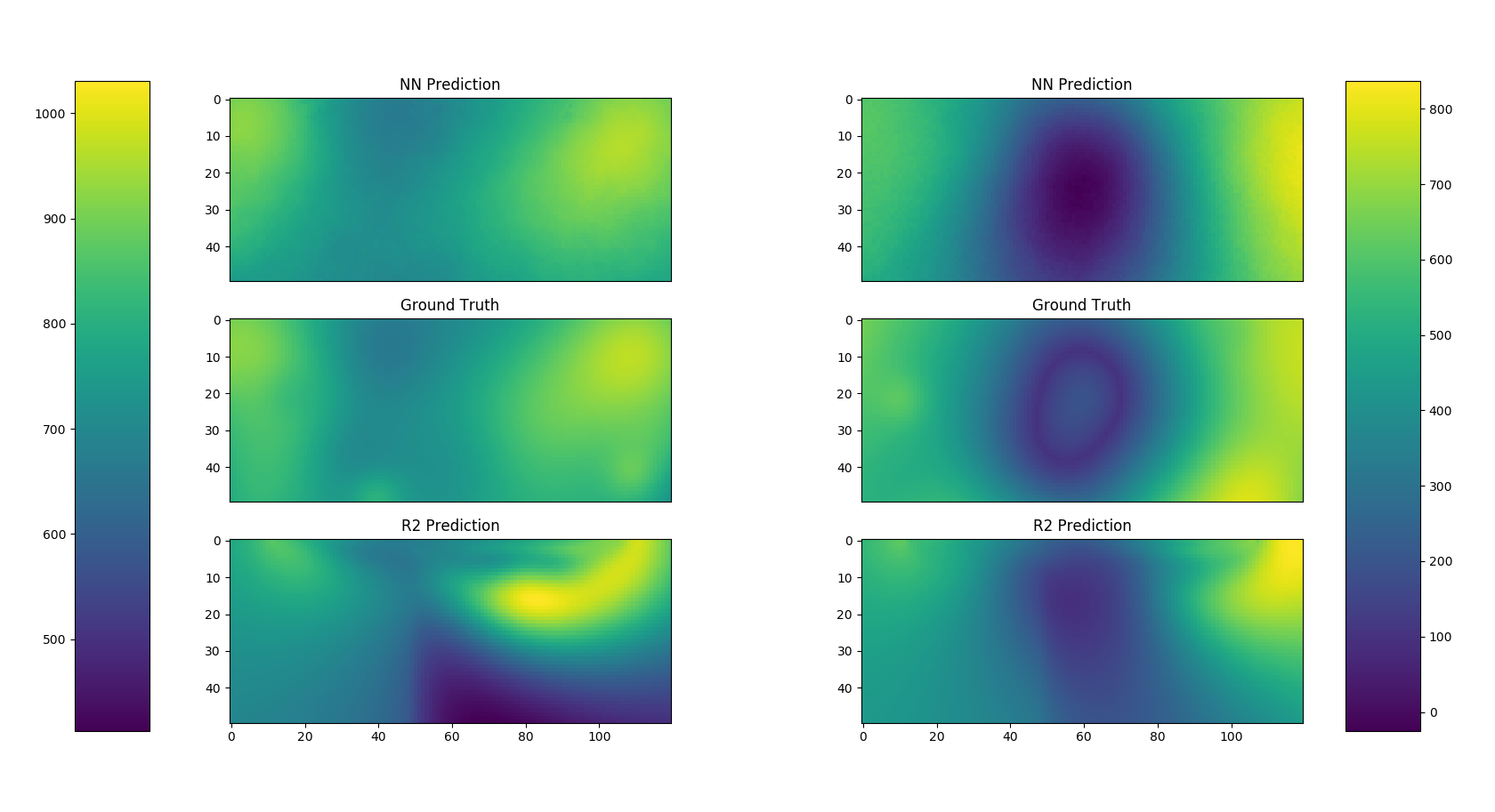} \\
		\hline \\
		\includegraphics[scale=0.4]{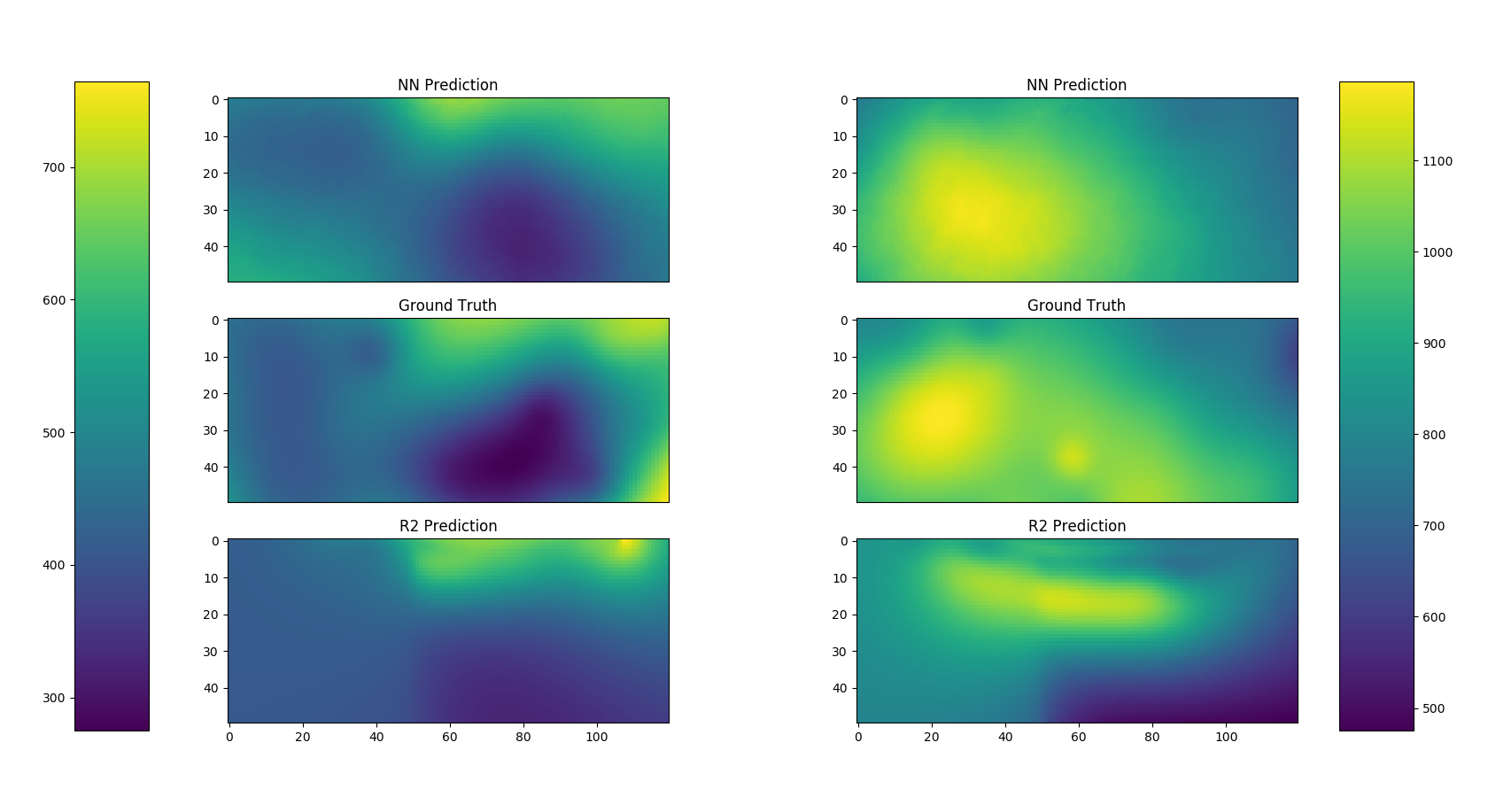}
	\end{tabular}
	\caption{Illustrative examples showing the predicted sub-surface resistivity profile. Within each triplet of images the middle image is the true ground, the upper one is the prediction using the neural network and the lower on is the prediction of R2 using the classical inversion method. All the images within the triplet use the same color scale. 
	\textbf{TOP LEFT}: 
	best example of the NN prediction. $L_1$ losses are: $L_1(NN)=12,\ L_1(R2)=260$.
	\textbf{TOP RIGHT}: 
	worst example of the NN prediction. $L_1$ losses are: $L_1(NN)=55,\ L_1(R2)=204$.
	\textbf{BOTTOM LEFT}: 
	best example of the R2 prediction. $L_1$ losses are: $L_1(NN)=16,\ L_1(R2)=46$.
	\textbf{BOTTOM RIGHT}: 
	worst example of the R2 prediction. $L_1$ losses are: $L_1(NN)=37,\ L_1(R2)=446$.
	}
	\label{examples}
\end{table}

The usual way to examine the results of an inversion process is by comparing the ERT measurements $(E)$ obtained from the forward solution of the true ground to the ERT measurements $(\widetilde{E})$ of the predicted ground and calculating some loss function $\mathcal{L}(E, \widetilde{E})$ . When the loss function $\mathcal{L}(E, \widetilde{E})$ is small enough the inversion process stops. However, this does not guaranty that the predicted ground $\widetilde{G}$ is similar to the true ground $G$. In the scenario of real measurements in the field, this is the best one can do with an inversion process. However, using simulated data we can make a comparison between the real goal of the inversion process - the predicted ground $\widetilde{G}$ and the true ground $G$.

In order to compare the results of the NN to the classical inversion method we used R2 program to predict the grounds for the 1500 test examples. The parameters we used for the inversion process were: background resistivity of $600 \Omega m$, tolerance of $0.1$, maximum of 10 iterations 
\footnote{For more detail about the meaning of each parameter, please refere to the R2 manual \cite{R2}}. 
The requested output from the inversion has the same size and resolution as the NN predictions and the true grounds: width of $120 m$ and depth of $50 m$ with resolution of $1m^2$. 

The comparison between the NN prediction and the predictions of the inversion process is done by using two parameters.
The first one is the $L_1$ loss \eqref{eq: L1} of each prediction made by NN and R2. 
It is also interesting to have a measure of the overall similarity of values between the prediction and the true ground. This can be done by using the same $L_1$ distance between the histogram of the predicted ground and the histogram of the true ground. This way, slightly spatially shifted features will not cause large penalty.

Figure \ref{fig: l1_and_l1h} depicts the histograms of the $L_1$ loss for the predictions made by both the NN and R2. It is very clear that while R2 have a very large span of errors, ranging from 50 to 150, the NN achieves errors of around 50 in its worst case. While $L_1$ Loss is very sensitive to the value of each $1m^2$ pixel in the ground, the $L_1$ Loss of the histogram is a comparison of the distributions. Comparing the values of $L_1$ for both histograms with respect to the true ground (right panel of Fig. \ref{fig: l1_and_l1h}) we find that the NN have a narrow distribution of $L_1$ Loss values and the values are closer to 0 than the $L_1$ Loss of the predictions made by R2.

An illustration of the different sub-surface reconstruction is shown in Table \ref{examples}. Within each column of 3 images, the middle panel is the true ground, the upper panel is the prediction of the neural network and the lower panel is the prediction of the inversion process using R2. It is clear that in all the examples the NN has a better performance than the classical inversion method.

\newpage
\section{Conclusions}\label{sec: Conclusions}
We have demonstrate the ability to reconstruct the sub-surface resistivity profile from an ERT survey using a trained neural network. 
As shown in Figure \ref{fig: l1_and_l1h}, the predictions using neural network are more accurate than the ground predicted by using the traditional inversion method. Moreover, the neural network is capable of predicting the underground resistivity structure even at depth of 40-50 meters which are well beyond the maximal depth of the described inversion method. Although the mapping at those depths is not perfect, the main features are visible and recognizable and in many practical cases a slight deformation or offset is within the tolerance of acceptable solutions.
Another feature of the neural network is that the prediction process is obviously faster than conventional inversion algorithms and results are produced within milliseconds of receiving the ERT measurements from the field. Future research will include the 
extension of this approach for using other survey protocols, reducing the number of measurements in the survey and taking into account the effect of various noises on the performance of the neural network predictions.

\bibliography{mlert}

\begin{thebibliography}{16}
\expandafter\ifx\csname natexlab\endcsname\relax\def\natexlab#1{#1}\fi
\expandafter\ifx\csname bibnamefont\endcsname\relax
  \def\bibnamefont#1{#1}\fi
\expandafter\ifx\csname bibfnamefont\endcsname\relax
  \def\bibfnamefont#1{#1}\fi
\expandafter\ifx\csname citenamefont\endcsname\relax
  \def\citenamefont#1{#1}\fi
\expandafter\ifx\csname url\endcsname\relax
  \def\url#1{\texttt{#1}}\fi
\expandafter\ifx\csname urlprefix\endcsname\relax\def\urlprefix{URL }\fi
\providecommand{\bibinfo}[2]{#2}
\providecommand{\eprint}[2][]{\url{#2}}

\bibitem[{\citenamefont{Morelli and LaBrecque}(1996)}]{morelli1996advances}
\bibinfo{author}{\bibfnamefont{G.}~\bibnamefont{Morelli}} \bibnamefont{and}
  \bibinfo{author}{\bibfnamefont{D.~J.} \bibnamefont{LaBrecque}},
  \bibinfo{journal}{European Journal of Environmental and Engineering
  Geophysics} \textbf{\bibinfo{volume}{1}}, \bibinfo{pages}{171}
  (\bibinfo{year}{1996}).

\bibitem[{\citenamefont{Saad et~al.}(2012)\citenamefont{Saad, Nawawi, and
  Mohamad}}]{saad2012groundwater}
\bibinfo{author}{\bibfnamefont{R.}~\bibnamefont{Saad}},
  \bibinfo{author}{\bibfnamefont{M.}~\bibnamefont{Nawawi}}, \bibnamefont{and}
  \bibinfo{author}{\bibfnamefont{E.}~\bibnamefont{Mohamad}},
  \bibinfo{journal}{Electronic Journal of Geotechnical Engineering}
  \textbf{\bibinfo{volume}{17}}, \bibinfo{pages}{369} (\bibinfo{year}{2012}).

\bibitem[{\citenamefont{Tsourlos and Ogilvy}(1999)}]{tsourlos1999algorithm}
\bibinfo{author}{\bibfnamefont{P.}~\bibnamefont{Tsourlos}} \bibnamefont{and}
  \bibinfo{author}{\bibfnamefont{R.}~\bibnamefont{Ogilvy}},
  \bibinfo{journal}{Journal of the Balkan Geophysical Society}
  \textbf{\bibinfo{volume}{2}}, \bibinfo{pages}{30} (\bibinfo{year}{1999}).

\bibitem[{\citenamefont{Camporese et~al.}(2015)\citenamefont{Camporese,
  Cassiani, Deiana, Salandin, and Binley}}]{WRCR21447}
\bibinfo{author}{\bibfnamefont{M.}~\bibnamefont{Camporese}},
  \bibinfo{author}{\bibfnamefont{G.}~\bibnamefont{Cassiani}},
  \bibinfo{author}{\bibfnamefont{R.}~\bibnamefont{Deiana}},
  \bibinfo{author}{\bibfnamefont{P.}~\bibnamefont{Salandin}}, \bibnamefont{and}
  \bibinfo{author}{\bibfnamefont{A.}~\bibnamefont{Binley}},
  \bibinfo{journal}{Water Resources Research} \textbf{\bibinfo{volume}{51}},
  \bibinfo{pages}{3277} (\bibinfo{year}{2015}), ISSN \bibinfo{issn}{1944-7973},
  \urlprefix\url{http://dx.doi.org/10.1002/2014WR016017}.

\bibitem[{\citenamefont{Vanderborght et~al.}(2005)\citenamefont{Vanderborght,
  Kemna, Hardelauf, and Vereecken}}]{WRCR10223}
\bibinfo{author}{\bibfnamefont{J.}~\bibnamefont{Vanderborght}},
  \bibinfo{author}{\bibfnamefont{A.}~\bibnamefont{Kemna}},
  \bibinfo{author}{\bibfnamefont{H.}~\bibnamefont{Hardelauf}},
  \bibnamefont{and}
  \bibinfo{author}{\bibfnamefont{H.}~\bibnamefont{Vereecken}},
  \bibinfo{journal}{Water Resources Research} \textbf{\bibinfo{volume}{41}},
  \bibinfo{pages}{n/a} (\bibinfo{year}{2005}), ISSN \bibinfo{issn}{1944-7973},
  \bibinfo{note}{w06013},
  \urlprefix\url{http://dx.doi.org/10.1029/2004WR003774}.

\bibitem[{\citenamefont{Loke et~al.}(2014)\citenamefont{Loke, Dahlin, and
  Rucker}}]{loke2014smoothness}
\bibinfo{author}{\bibfnamefont{M.}~\bibnamefont{Loke}},
  \bibinfo{author}{\bibfnamefont{T.}~\bibnamefont{Dahlin}}, \bibnamefont{and}
  \bibinfo{author}{\bibfnamefont{D.}~\bibnamefont{Rucker}}
  (\bibinfo{year}{2014}).

\bibitem[{\citenamefont{Wu et~al.}(2016)\citenamefont{Wu, Schuster, Chen, Le,
  Norouzi, Macherey, Krikun, Cao, Gao, Macherey et~al.}}]{NN-Translate}
\bibinfo{author}{\bibfnamefont{Y.}~\bibnamefont{Wu}},
  \bibinfo{author}{\bibfnamefont{M.}~\bibnamefont{Schuster}},
  \bibinfo{author}{\bibfnamefont{Z.}~\bibnamefont{Chen}},
  \bibinfo{author}{\bibfnamefont{Q.~V.} \bibnamefont{Le}},
  \bibinfo{author}{\bibfnamefont{M.}~\bibnamefont{Norouzi}},
  \bibinfo{author}{\bibfnamefont{W.}~\bibnamefont{Macherey}},
  \bibinfo{author}{\bibfnamefont{M.}~\bibnamefont{Krikun}},
  \bibinfo{author}{\bibfnamefont{Y.}~\bibnamefont{Cao}},
  \bibinfo{author}{\bibfnamefont{Q.}~\bibnamefont{Gao}},
  \bibinfo{author}{\bibfnamefont{K.}~\bibnamefont{Macherey}},
  \bibnamefont{et~al.}, \bibinfo{journal}{CoRR}
  \textbf{\bibinfo{volume}{abs/1609.08144}} (\bibinfo{year}{2016}),
  \eprint{1609.08144}, \urlprefix\url{http://arxiv.org/abs/1609.08144}.

\bibitem[{\citenamefont{Zhu et~al.}(2017)\citenamefont{Zhu, Park, Isola, and
  Efros}}]{NN-GAN}
\bibinfo{author}{\bibfnamefont{J.-Y.} \bibnamefont{Zhu}},
  \bibinfo{author}{\bibfnamefont{T.}~\bibnamefont{Park}},
  \bibinfo{author}{\bibfnamefont{P.}~\bibnamefont{Isola}}, \bibnamefont{and}
  \bibinfo{author}{\bibfnamefont{A.~A.} \bibnamefont{Efros}},
  \bibinfo{journal}{arXiv preprint arXiv:1703.10593}  (\bibinfo{year}{2017}).

\bibitem[{\citenamefont{Naparstek and Cohen}(2017)}]{NN-Oshri}
\bibinfo{author}{\bibfnamefont{O.}~\bibnamefont{Naparstek}} \bibnamefont{and}
  \bibinfo{author}{\bibfnamefont{K.}~\bibnamefont{Cohen}},
  \bibinfo{journal}{CoRR} \textbf{\bibinfo{volume}{abs/1704.02613}}
  (\bibinfo{year}{2017}), \eprint{1704.02613},
  \urlprefix\url{http://arxiv.org/abs/1704.02613}.

\bibitem[{\citenamefont{Raiche}(1991)}]{NN1991}
\bibinfo{author}{\bibfnamefont{A.}~\bibnamefont{Raiche}},
  \bibinfo{journal}{Geophys. J. Int.} \textbf{\bibinfo{volume}{105}},
  \bibinfo{pages}{629} (\bibinfo{year}{1991}).

\bibitem[{\citenamefont{Alireza~HAJIAN}(2012)}]{NN-gravity}
\bibinfo{author}{\bibfnamefont{P.~S.} \bibnamefont{Alireza~HAJIAN},
  \bibfnamefont{Hossein~ZOMORRODIAN}}, \bibinfo{journal}{Acta Geophysica}
  \textbf{\bibinfo{volume}{60}}, \bibinfo{pages}{1043} (\bibinfo{year}{2012}).

\bibitem[{\citenamefont{Yuan et~al.}(2018)\citenamefont{Yuan, Liu, Wang, Wang,
  and Shi}}]{NN-seismic}
\bibinfo{author}{\bibfnamefont{S.}~\bibnamefont{Yuan}},
  \bibinfo{author}{\bibfnamefont{J.}~\bibnamefont{Liu}},
  \bibinfo{author}{\bibfnamefont{S.}~\bibnamefont{Wang}},
  \bibinfo{author}{\bibfnamefont{T.}~\bibnamefont{Wang}}, \bibnamefont{and}
  \bibinfo{author}{\bibfnamefont{P.}~\bibnamefont{Shi}}, \bibinfo{journal}{IEEE
  Geoscience and Remote Sensing Letters} \textbf{\bibinfo{volume}{15}},
  \bibinfo{pages}{272} (\bibinfo{year}{2018}), ISSN \bibinfo{issn}{1545-598X}.

\bibitem[{\citenamefont{Ehret}(2010)}]{NN-GPR}
\bibinfo{author}{\bibfnamefont{B.}~\bibnamefont{Ehret}},
  \bibinfo{journal}{Geoderma} \textbf{\bibinfo{volume}{160}},
  \bibinfo{pages}{111} (\bibinfo{year}{2010}).

\bibitem[{\citenamefont{Upendra~K.Singh}(2013)}]{NN-VES}
\bibinfo{author}{\bibfnamefont{S.}~\bibnamefont{Upendra~K.Singh},
  \bibfnamefont{R.K.Tiwari}}, \bibinfo{journal}{Computers \& Geosciences}
  \textbf{\bibinfo{volume}{52}}, \bibinfo{pages}{246} (\bibinfo{year}{2013}).

\bibitem[{\citenamefont{Loke}(2006)}]{loke-tutorial}
\bibinfo{author}{\bibfnamefont{M.}~\bibnamefont{Loke}},
  \bibinfo{journal}{Geotomo Software, Malaysia}  (\bibinfo{year}{2006}),
  \urlprefix\url{http://www.geotomosoft.com/coursenotes.zip}.

\bibitem[{\citenamefont{Binley}()}]{R2}
\bibinfo{author}{\bibfnamefont{A.}~\bibnamefont{Binley}},
  \emph{\bibinfo{title}{R2 version 3.1 june 2016}},
  \urlprefix\url{http://www.es.lancs.ac.uk/people/amb/Freeware/R2/R2.htm}.

\end{thebibliography}

\end{document}